\documentclass[fleqn,usenatbib]{mnras}

\usepackage{newtxtext,newtxmath}

\usepackage[T1]{fontenc}
\newcommand{\St}{\mathrm{St}}

\DeclareRobustCommand{\VAN}[3]{#2}
\let\VANthebibliography\thebibliography
\def\thebibliography{\DeclareRobustCommand{\VAN}[3]{##3}\VANthebibliography}

\usepackage{graphicx}	
\usepackage{amsmath}	
\usepackage{comment}
\usepackage{CJK}

\newcommand{\ctext}[1]{\textup{\begin{CJK*}{UTF8}{bkai}#1\ignorespacesafterend\end{CJK*}}}

\title[Fragmentation inside spiral waves]{Particle fragmentation inside planet-induced spiral waves}

\author[L.E.J. Eriksson et al.]{
Linn E.J. Eriksson,$^{1}$\thanks{E-mail: linn.eriksson@stonybrook.edu}
Chao-Chin Yang (\ctext{楊朝欽}),$^{2}$
Philip J. Armitage$^{3,4}$ 
\\
$^{1}$ Institute for Advanced Computational Sciences, Stony Brook University, Stony Brook, NY, 11794-5250, USA \\
$^{2}$ Department of Physics and Astronomy, The University of Alabama, Box 870324, Tuscaloosa, AL 35487-0324, USA \\
$^{3}$ Center for Computational Astrophysics, Flatiron Institute, 162 Fifth Avenue, New York, NY 10010, USA \\
$^{4}$ Department of Physics and Astronomy, Stony Brook University, Stony Brook, NY 11794, USA
}

\date{Accepted XXX. Received YYY; in original form ZZZ}

\pubyear{2024}

\begin{document}
\label{firstpage}
\pagerange{\pageref{firstpage}--\pageref{lastpage}}
\maketitle

\begin{abstract}
Growing planets interact with their surrounding protoplanetary disk, generating feedback effects that may promote or suppress nearby planet formation. We study how spiral waves launched by planets affect the motion and collisional evolution of particles in the disk. To this end, we perform local 2D hydrodynamical simulations that include a gap-opening planet and integrate particle trajectories within the gas field. Our results show that particle trajectories bend at the location of the spiral wave, and collisions occurring within the spiral exhibit significantly enhanced collisional velocities compared to elsewhere. To quantify this effect, we ran simulations with varying planetary masses and particle sizes. The resulting collisional velocities within the spiral far exceed the typical fragmentation threshold, even for collisions between particles of relatively similar sizes and for planetary masses below the pebble isolation mass. If collisions within the spiral are frequent, this effect could lead to progressively smaller particle sizes as the radial distance from the planet decreases, impacting processes such as gap filtering, pebble accretion, and planetesimal formation. 

\end{abstract}




\section{Introduction}
Growing planets launch spiral waves in the protoplanetary disk, the prominence of which depends predominantly on the mass of the planet, the thermodynamics, and the viscosity of the gas \citep{Paardekooper2023}. In the case of massive planets, the torque exerted on the disk drives gas away from the planetary orbit, resulting in the opening of a planetary gap. The pressure maximum that forms at the gap edge can act to trap dust and is often invoked to explain the presence of gaps and rings in observations of protoplanetary disks (e.g. \citealt{Dipierro2015, Zhang2018}).

As the disk evolves, initially micron-sized dust grains collide and grow to form larger particles called pebbles. Further collisional growth is impaired by radial drift, fragmentation and bouncing \citep{Birnstiel2016}. Radial drift occurs because gas in the disk orbits at a sub-Keplerian velocity, causing pebbles that move at closer to Keplerian velocity to experience a headwind, lose angular momentum, and drift towards the star \citep{Adachi1976,Weidenschilling1977}. Particles are also subject to turbulent mixing \citep{Youdin2007}. The resulting velocities from all of these effects depend on how well coupled the particles are to the gas.

The relative velocity between pairs of particles increases as the difference in particle size becomes larger. High-velocity collisions tend to result in fragmentation rather than sticking, with collision experiments finding fragmentation velocities, $v_{\rm frag}$, of about $1-10\, \textrm{m}\, \textrm{s}^{-1}$ \citep{BlumWurm2008,GundlachBlum2015,MusiolikWurm2019,Musiolik2021}. \citet{Drazkowska2019} found that efficient fragmentation among particles trapped at the planetary gap edge leads to an enhancement of small particles, which has important implications for the filtering of particles across planetary gaps. In this work, we study whether fragmentation can also be enhanced inside the planet-induced spiral wave. To this end, we perform local 2D hydrodynamical simulations of planet-disk interactions and integrate the orbits of particles on top of the gas field.

\section{NUMERICAL MODELLING}\label{sect: model}
We adopt the local-shearing-sheet approximation \citep{GoldreichLyndenBell1965} to model the evolution of a razor-thin protoplanetary disk perturbed by a planet. Our methodology closely follows that of \citet{YangZhu2020} for the limit of no dust back-reaction, with the exception of added viscosity to allow for gas-gap equilibrium. Below is a summary of our model; for a detailed description, we refer to \citet{YangZhu2020}.

\subsection{Gas dynamics}
The protoplanetary disk considered in this work is isothermal, non-self-gravitating, non-magnetized and razor-thin. We model a local patch of the disk located at an arbitrary radial distance $r=r_0$ from the star, which rotates around the star with the Keplerian angular frequency $\Omega_{\rm K}$. The shearing sheet is mapped by a fixed regular grid with coordinates $x=r-r_0$ and $y=r(\theta-\theta_0)$, where $\theta_0$ is an arbitrary reference azimuthal angle. All velocities are calculated with respect to the background Keplerian shear velocity $-3\Omega_{\rm K}x\hat{\bf{e}}_y/2$. The gaseous component is evolved with the PENCIL CODE \citep{BrandenburgDobler2002,Pencil2021}, which employs sixth-order spatial derivatives and a third-order Runge-Kutta method for time integration.  

We impose a constant radial pressure gradient such that $\Delta u / c_{\rm s} = 0.05$, where $\Delta u$ is the reduction in azimuthal gas velocity compared to the Keplerian velocity, and $c_{\rm s}$ is the sound speed. The formulation and implementation of the viscous stress tensor is described by Eqs. (9) and (12) in \citet{Brandenburg2003}, and we analytically add the background shear. We approximate the viscosity using the dimensionless alpha approach, $\nu = \alpha \Omega_{\rm K} H_{\rm g}^2$ \citep{ShakuraSunyaev1973}. Here, $H_{\rm g}=c_{\rm s}/\Omega_{\rm K}$ is the vertical gas scale height, and $\alpha$ governs the efficiency of turbulent transport. We use a constant value of $\alpha=0.01$ throughout this work. 

The planet is placed at the centre of the shearing sheet and modelled by a smoothed gravitational potential,
\begin{equation}
    \Phi(d) = -GM_{\rm p} \frac{d^2+3r_{\rm s}^2/2}{(d^2+r_{\rm s}^2)^{3/2}},
\end{equation}
where $G$ is the gravitational constant, $M_{\rm p}$ is the planetary mass, $d=\sqrt{x^2+y^2}$ is the distance to the planet, and $r_{\rm s}$ is the smoothing length \citep{Dong2011,Zhu2012}. The planetary mass is defined in terms of the thermal mass, 
\begin{equation}
    M_{\rm th}=\frac{c_{\rm s}^3}{G\Omega_{\rm K}},
\end{equation}
and we adopt a smoothing length of $r_{\rm s}=0.8R_{\rm H}$, where
\begin{equation}
   R_{\rm H}=r_0\left(\frac{M_{\rm p}}{3M_*}\right)^{1/3}=H_{\rm g}\left(\frac{M_{\rm p}}{3M_{\rm th}}\right)^{1/3}
\end{equation}
is the Hill radius. The gas is initialized in equilibrium state with a constant surface density $\Sigma_0$; in other words, the global radial pressure gradient is imposed as an external background. The gravitational force from the planet is gradually increased from 0 to its full magnitude over $10P$ by varying $G$, where $P=2\pi/\Omega_{\rm K}$ is the orbital period. 

We adopt a large computational domain of $16H_{\rm g}$ in the radial dimension and $32H_{\rm g}$ in the azimuthal dimension, with a fixed resolution of 32 cells per $H_{\rm g}$. Standard sheared-periodic boundary conditions are employed \citep{Brandenburg1995, Hawley1995}, along with a damping zone of width $1H_{\rm g}$ near the radial boundary to damp waves generated by the planet (eq. 4-6 in \citealt{YangZhu2020}). We perform simulations with $M_{\rm p}/M_{\rm th} = 0.25$, $0.5$, $0.75$ and $1$. Each system is evolved up to $t=200P$, by which point all systems have long reached equilibrium. Our gap depths agree within 10\% of the fit provided by \citet{Kanagawa2015}.

\subsection{Particle dynamics}
We use the equilibrium state of the gas disk at $t=200P$ to post-process the simulations and track the movement of particles. The particles are modelled as Lagrangian particles, with Stokes number $\St=\Omega_{\rm K}t_{\rm s}$, where $t_{\rm s}$ is the characteristic time-scale for reducing the relative velocity between a particle and its surrounding gas by the drag force. The Stokes number predominantly depends on particle size and the local gas density, with the peak drift velocity occurring for particles with $\St=1$. In the PENCIL CODE simulations of \citet{YangZhu2020}, the Stokes number was kept constant as the particles moved through the gas disk. In our post-processing routine, we take the dependence of gas density into account by using: $t_{\rm s} = t_{\rm s,0} / (\Sigma_{\rm g}/\Sigma_0)$ \citep{Weidenschilling1977}. The equations of motion for each particle are:
\begin{equation}
    \frac{d\bf{x}_{\rm p}}{dt} = {\bf v_{\rm p}}-\frac{3}{2}\Omega_{\rm K}x_{\rm p}\hat{\bf{y}},
\end{equation}
\begin{equation}
    \frac{d\bf{v}_{\rm p}}{dt} = \left(2\Omega_{\rm K}v_{p,y}\hat{\bf{x}} - \frac{1}{2}\Omega_{\rm K}v_{p,x}\hat{\bf{y}} \right) + \frac{\bf{u}-\bf{v_p}}{t_{\rm s}} - \nabla\Phi,
\end{equation}
where $\bf{x}_{\rm p}$ is the particle position, and $\bf{v}_{\rm p}$ is the particle velocity relative to the background shear flow.

We integrate the equations of motion using a fourth-order Runge-Kutta method and apply bilinear interpolation to determine the gas velocity $\bf{u}$ and $\Sigma_{\rm g}$ at the particle positions. The particles are introduced just interior of the radial damping zone, with random azimuthal positions and initial drift velocities as described by \citet{Adachi1976} and \citet{Weidenschilling1977}. We consider particles with $\St=0.1$, $0.075$, $0.05$, $0.025$, $0.01$, and as discussed in Section~\ref{sect: results}, we only integrate one trajectory for each St. The particle trajectories are integrated until one of the following occurs: (a) the particle passes through the gap; (b) the particle enters a region within $0.2R_{\rm H}$ of the planet; or (c) the particle's drift halts at the gap edge. Since the background gas field is constant and there is no random particle motion, the particle trajectories are deterministic. We compared post-processed results using constant Stokes number against particle trajectories evolved with the Pencil Code and found excellent agreement. We also confirmed that the radial drift velocity far from the planet agrees with the analytical drift rate assuming a zero radial gas velocity.

\section{Simulation results}\label{sect: results}

\begin{figure*}
    \centering
    \includegraphics[width=1.63\columnwidth]{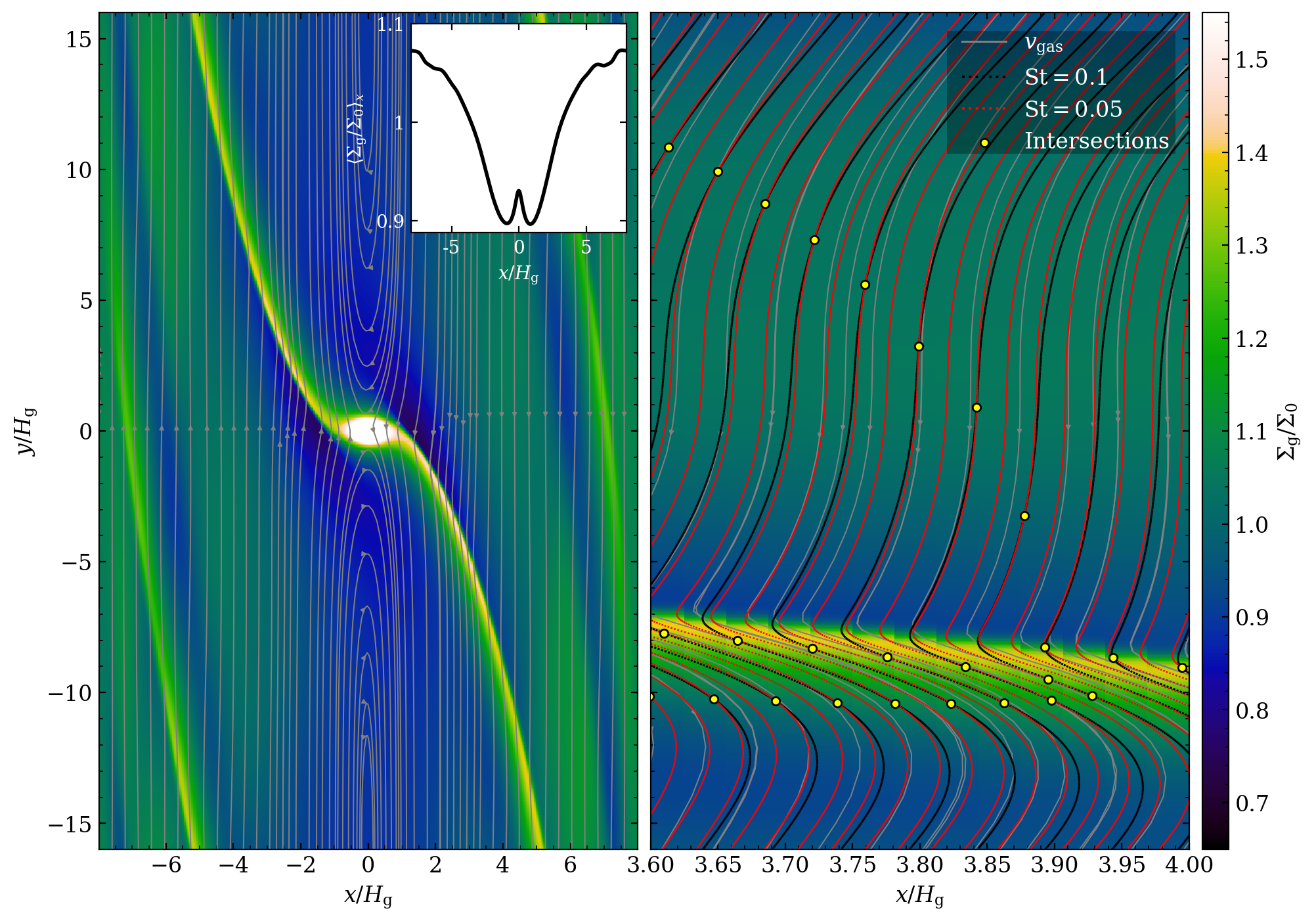}
    \caption{Left: Gas surface density at $t=200\, P$ from our simulation with $M_{\rm p}/M_{\rm th} = 0.5$, with gas streamlines overlaid. The azimuthally averaged gas surface density profile is shown in the top right corner. Right: Same as the left panel, but zoomed in on a small radial portion of the domain. Trajectories for particles with $\St=0.1$ and $\St=0.05$ are shown with black and red lines, respectively. Scatter points mark the locations where the trajectories intersect. Gas streamlines and particle trajectories bend at the location of the spiral wave. }
    \label{fig:yVSx_rho}
\end{figure*}

\begin{figure*}
    \centering
    \includegraphics[width=2\columnwidth]{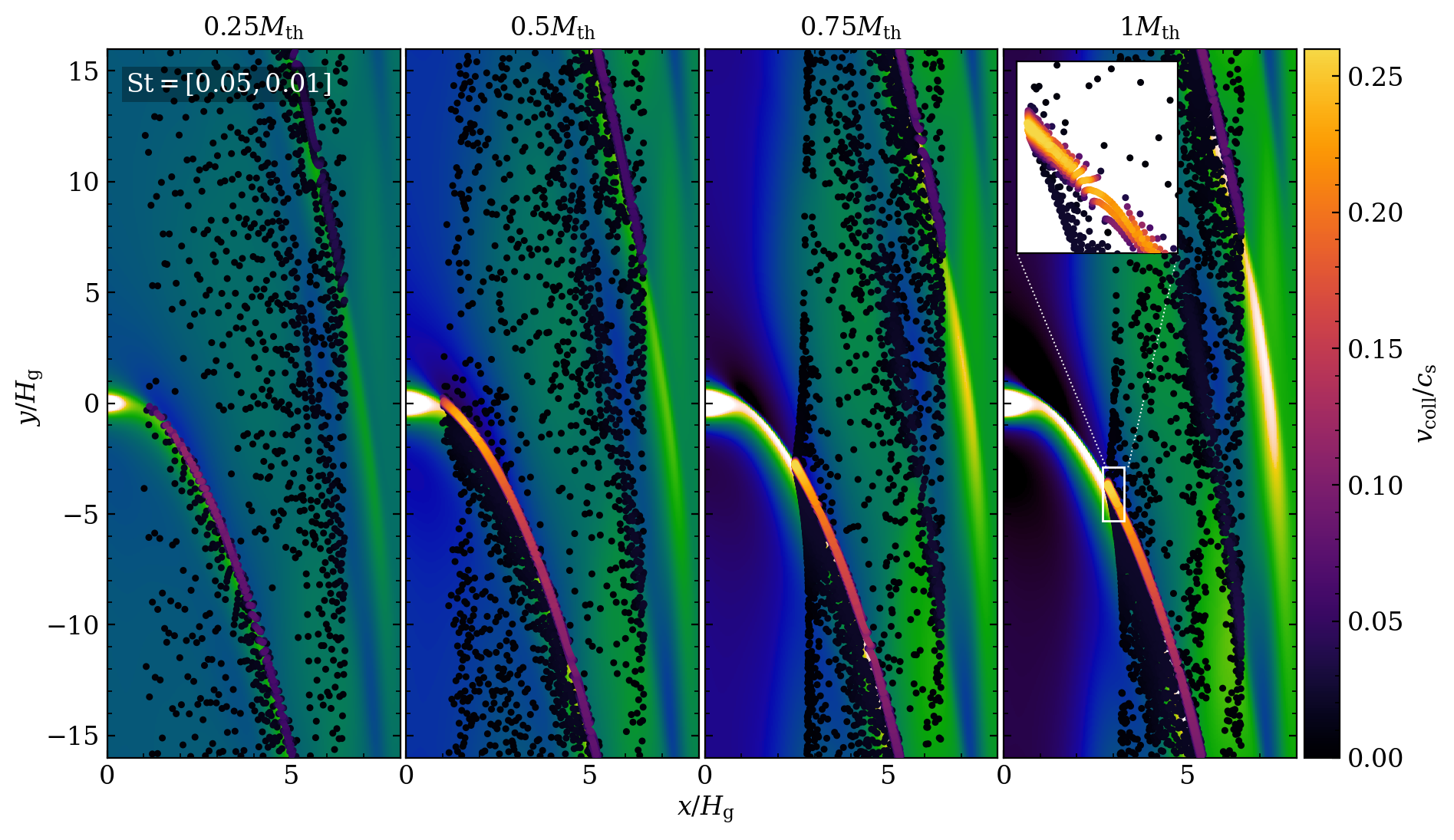}
    \caption{The scatter points show the locations and collisional velocities at all intersection points between the trajectories of particles with $\St=0.05$ and $\St=0.01$, for all considered planetary masses. The gas surface density is shown in the background, using the same color scale as in Fig.~\ref{fig:yVSx_rho}. Collisional velocities are significantly enhanced for collisions occurring inside the spiral wave. The collisional velocity increases with planetary mass, decreases with radial distance from the planet, and decreases with distance from the center of the spiral wave.  }
    \label{fig:yVSx_rho_Vcol}
\end{figure*}

Fig.~\ref{fig:yVSx_rho} shows the gas surface density, gas streamlines, and trajectories of particles with $\St=0.1$ and $\St=0.05$ from our simulation with $M_{\rm p}/M_{\rm th} = 0.5$ (see Fig.~\ref{fig:yVSx_ux} in the Appendix for color maps of the gas velocity field). The planet induces a pair of spiral arms that wrap around the azimuthal periodic boundary and propagate radially away from the planet until they reach the radial damping zone, where they are damped. The planet exerts a torque on the gas, resulting in gap formation. Simultaneously, the viscous force acts to refill the gap, eventually leading to an equilibrium state. Gas located deep within the gap executes horseshoe orbits relative to the planet. Further away, the streamlines are relatively straight but bend at the location of the spiral wave. This bending of the streamlines decreases radially away from the planet, as the spiral arms become less prominent. The right panel of Fig.~\ref{fig:yVSx_rho} shows how both the gas streamlines and particle trajectories bend at the location of the spiral wave. The particle trajectories bend less than the gas streamlines, and particles with larger $\St$ bend less than those with smaller $\St$, which are more closely coupled to the gas. 

Given that our particle trajectories are deterministic, and since we do not aim to constrain collisional probabilities in this work, we limit our study to one integrated particle trajectory per combination of $\St$ and $M_{\rm p}$. To study particle collisions, we find all intersections between trajectories of particles with different $\St$ and use linear interpolation in $x$ and $y$ to obtain the velocities at the intersection points. This approach ignores the time dimension; rather, it assumes a constant stream of particles along the trajectories. In the right panel of Fig.~\ref{fig:yVSx_rho}, we highlight intersections between trajectories of particles with $\St=0.1$ and $\St=0.05$. The relative velocity at the intersection point, henceforth referred to as the collisional velocity, $v_{\rm coll}$, is obtained as:
\begin{equation}
    v_{\rm coll} = \sqrt{\Delta v_{p,x}^2 + \Delta v_{p,y}^2},
\end{equation}
where $\Delta v_{p,x/y}$ is the x/y-component of particle velocity, respectively. If the calculated collisional velocity exceeds the fragmentation velocity, it implies that if the two particles were to collide at that location, they would likely fragment (see Section~\ref{sect:conclusions} for a discussion on mass transfer).

Fig.~\ref{fig:yVSx_rho_Vcol} shows the locations of all intersections between our trajectories of particles with $\St=0.05$ and $\St=0.01$, along with the corresponding collisional velocities, for all four planetary masses considered. Collisions occurring inside the spiral wave exhibit significantly higher collisional velocities than those between the spirals. The collisional velocity within the spiral decreases both radially away from the planet and with distance from the center of the spiral. Collisional velocities increase with planetary mass, and for the highest planetary mass considered, we obtain velocities exceeding 20\% of the sound speed near the gap edge. Efficient fragmentation at the gap edge was also reported by \citet{Drazkowska2019}, who used a 2D coagulation model to study dust evolution in the presence of a Jupiter-mass planet. 

\begin{figure}
    \centering
    \includegraphics[width=1\columnwidth]{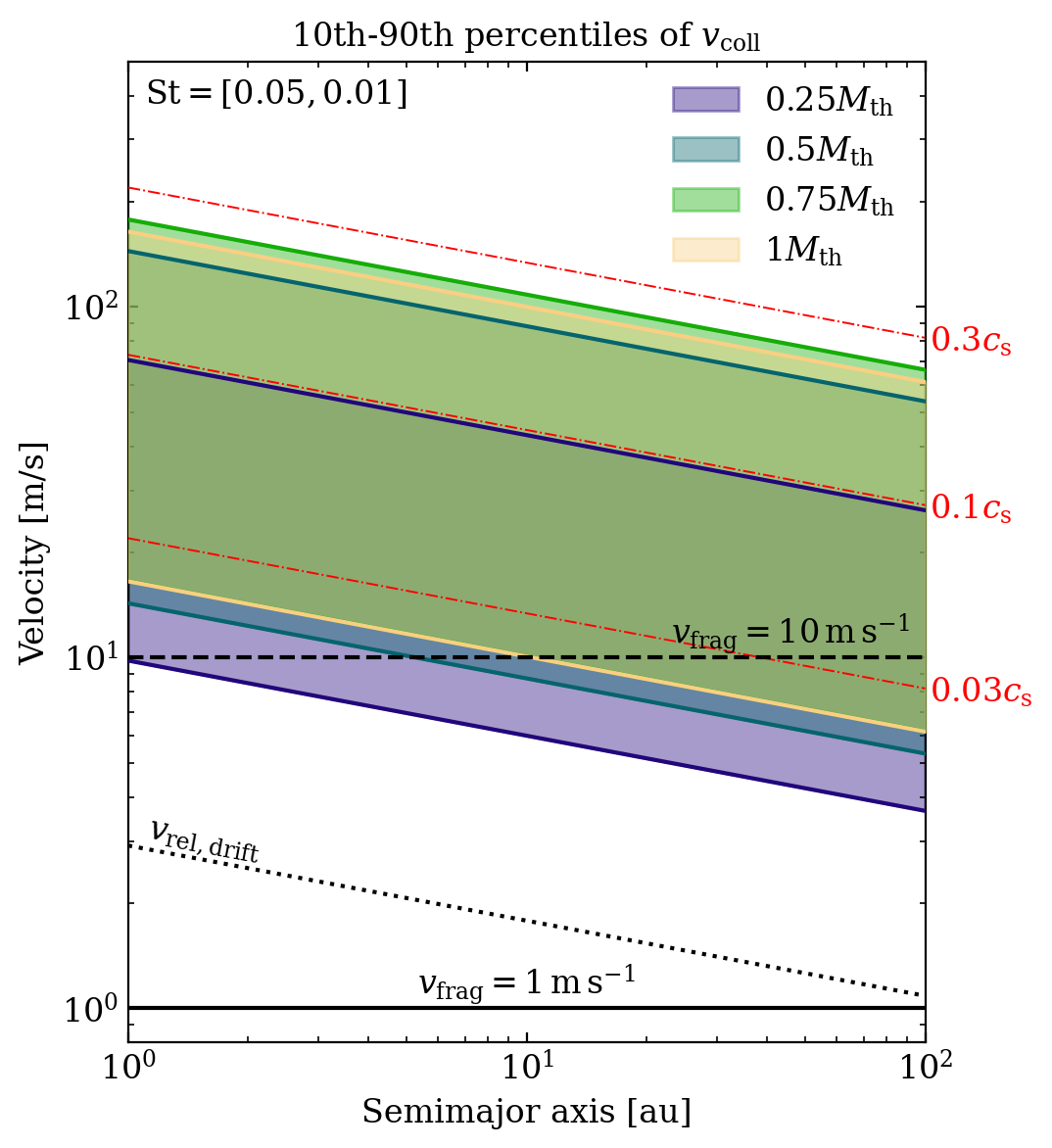}
    \caption{The colored areas indicate the region between the 10th and 90th percentiles of the same collision velocities shown in Fig.~\ref{fig:yVSx_rho_Vcol}, converted to SI units by assuming the sound speed profile by \citet{ChiangGoldreich1997}, but limited to collisions occurring within $y=\pm 0.5\, H_{\rm g}$ of the spiral center. Overlaid are the typically assumed fragmentation velocities and the differential drift velocity as a function of semimajor axis.}
    \label{fig:velocityVSr}
\end{figure}

In Fig.~\ref{fig:velocityVSr}, we compare the collisional velocities within the spiral wave to other relevant velocities, again for collisions occurring between particles of $\St=0.05$ and $\St=0.01$. The shown differential drift velocity, $v_{\rm rel,drift}$, is the difference in radial drift velocity for particles of $\St=0.05$ and $\St=0.01$ \citep{Weidenschilling1977}. We convert to SI units by multiplying by the sound speed, which is calculated assuming a temperature profile of $T = 150\, \textrm{K}\times (r/\textrm{au})^{-3/7}$ \citep{ChiangGoldreich1997} and a mean molecular weight of 2.34 \citep{Hayashi1981}. The majority of collisions occurring within the spiral wave happen at velocities well above the typically assumed fragmentation threshold.

\begin{figure*}
    \centering
    \includegraphics[width=2\columnwidth]{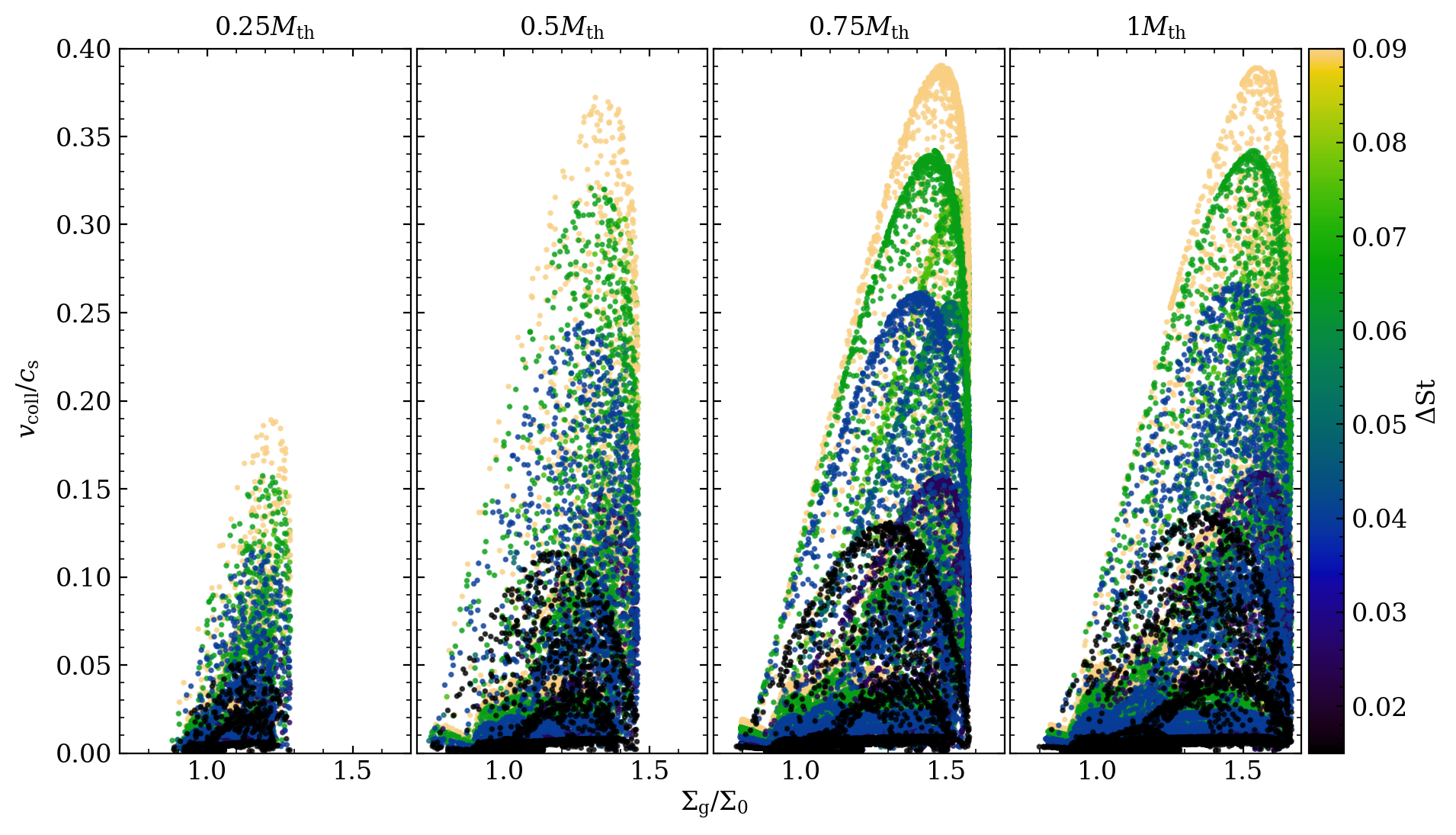}
    \caption{Collisional velocity as a function of local gas surface density for collisions between all particle pairs with different $\St$ considered in this work. The colors represent the difference in $\St$ between the colliding particles. The collisional velocity increases with $\Delta \St$, local gas surface density, and planetary mass, up to the gap-opening mass.}
    \label{fig:VcollVSrho}
\end{figure*}

Fig.~\ref{fig:VcollVSrho} shows the collisional velocity as a function of the local gas density at the intersection points, with the color of the scatter points indicating the difference in $\St$ between the colliding particles. The gas density enhancement inside the spiral decreases radially away from the planet and with distance away from the spiral center, making it a useful tracer of the spiral's strength. The collisional velocity increases with both local gas density and planetary mass, following the same trend observed in Fig.~\ref{fig:yVSx_rho_Vcol}. As expected, the collisional velocity also increases with $\Delta \St$, since particles with very different $\St$ follow very different trajectories as they cross the spiral wave, resulting in larger relative velocities. Collisions between particles near the gap edge can result in collisional velocities exceeding 35\% of the sound speed when large planetary masses are considered. Even for $\Delta \St = 0.025$, collisional velocities exceeding 5-10\% of the sound speed are commonly obtained, which can still be above the fragmentation limit in most regions of the disk (see Fig.~\ref{fig:velocityVSr}). The same analysis performed for simulations using constant $\St$ resulted in only up to $\sim$ 20\% higher collision velocities, implying that the gas velocity field and the bending of particle trajectories are the dominant cause for the enhanced collision velocities within the spiral, with non-axisymmetric density variations secondary.

\section{Concluding remarks}\label{sect:conclusions}

We have studied the motion of particles in the presence of embedded and gap-opening planets. Our results show that particle trajectories bend at the location of the planet-induced spiral wave. We integrated the trajectories of particles with different $\St$ and identified all intersections between particles with different $\St$, without accounting for the timing of the collisions. The collisional velocity at these intersection points was calculated, revealing significantly higher collisional velocities within the spiral compared to elsewhere. The collisional velocity within the spiral increases with planetary mass and $\Delta \St$ of the colliding particles, and decreases with both radial distance from the planet and distance from the center of the spiral wave. The velocities obtained predominantly exceed the fragmentation velocity, suggesting that collisions occurring within the spiral likely would result in fragmentation.

In our discussion so far, we have assumed that collisions occurring at velocities above the fragmentation threshold result in mass loss. This is the common assumption utilized in most studies and simulations of dust growth; however, laboratory experiments find that such collisions can result in a mass gain of the larger particle if the smaller particle is small enough (see \citealt{Johansen2014} and references therein). The efficiency of this mass transfer decreases with increasing collision velocity \citep{Meisner2013}, suggesting that the overall efficiency could be quite low for many of the high-velocity collisions considered in this study. It is also unclear what size difference that is required to be in the mass transfer regime, and the pairs of collisions considered in this work may not be over this threshold size difference, especially the small $\Delta \St$ pairs.

Coagulation typically results in peaked size distributions where most of the mass is concentrated in the largest particles, meaning that most of the mass should be in particles with similar $\St$ \citep{Birnstiel2012}. We find collisional velocities exceeding 5-10\% of the sound speed, corresponding to velocities above the typical fragmentation limit, even for $\Delta \St = 0.025$. For comparison, the relative velocity expected from collisions between similarly sized particles in turbulence is:
\begin{equation}\label{eq: vrel}
    v_{\rm rel, turb} = \sqrt{3\alpha_{\rm turb} \St}c_{\rm s}
\end{equation}
\citep{OrmelCuzzi2007}. Observations of protoplanetary disks generally suggest turbulent parameters on the order of $\alpha_{\rm turb} \sim 10^{-5}-10^{-3}$ \citep{Pinte2023}, which corresponds to collisional velocities of at most a few percent of the sound speed.

The impact of our results on dust evolution depends on the frequency of collisions within the spiral wave and whether the timescale for re-coagulation is faster or slower than the time between spiral wave passages (i.e., the orbital period). The dust growth timescale can be estimated as:
\begin{equation}
    \tau_{\rm grow} \simeq \frac{1}{\Omega_{\rm K}\Sigma_{\rm dust}/\Sigma_{\rm g}},
\end{equation}
for similarly sized particles and relative velocities given by Eq.~\ref{eq: vrel} \citep{Birnstiel2012}. For a typical dust-to-gas ratio of 1\%, this corresponds to a growth timescale that is much longer than the orbital period.

The model we present does not provide information on the frequency of collisions within the spiral wave. If collisions are common, our results suggest an increased amount of fragmentation, leading to progressively smaller particle sizes as the radial distance from the planet decreases. Smaller particles can cross the planetary gap more easily, as shown by \citet{Drazkowska2019}. Our findings thus support the idea that planetary gaps are leaky \citep{Stammler2023}, and may challenge the idea that the isotope dichotomy between carbonaceous (CC) and non-carbonaceous (NC) meteorites arose due to Jupiter's core forming early and separating the inner and outer Solar System into two separate reservoirs \citep{Kruijer2017} .  Additionally, smaller particles have a lower drift velocity, which could result in a buildup of particles beyond the planet's orbit, starting already beyond the gap edge. Furthermore, small particles are less efficiently accreted onto planets via pebble accretion \citep{Johansen2017} and are less efficient in forming planetesimals via the streaming instability \citep{Yang2017,LiYoudin2021}. 

The study that is presented is idealized and does not account for effects such as turbulence, dust-back reaction, dust coagulation or collisional probabilities. Our aim is to demonstrate the potential importance of particle fragmentation within planet-induced spiral waves. Additionally, our study is performed in 2D, neglecting any 3D effects. Spiral waves are in general weaker in 3D than in 2D, and the density contrast can be reduced by about a half \citep{Tanaka2002}. On the other hand, it remains unclear if the velocity perturbation in 3D is reduced by as much, and it can remain high in terms of the speed of sound \citep{Zhu2015,RabagoZhu2021}. As we have determined that the major driver of particle collisions is the velocity field in the gas instead of the density field, the collision speeds may remain high and over fragmentation threshold in 3D disks. To address the importance of our results for dust evolution and planet formation, future studies addressing all of the above considerations are necessary. 

\section*{Acknowledgements}
The authors wish to thank the anonymous referee for providing useful comments that led to an improved quality of the manuscript. The authors further wish to thank Anders Johansen and Zhaohuan Zhu for comments and advice regarding the model.
LE acknowledges funding by the Institute for Advanced Computational Science Postdoctoral Fellowship. 
CCY acknowledges the support from NASA via the Emerging Worlds program (\#80NSSC23K0653), the Astrophysics Theory Program (grant \#80NSSC24K0133), and the Theoretical and Computational Astrophysical Networks (grant \#80NSSC21K0497). PJA acknowledges support from NASA TCAN award 80NSSC19K0639, and from award 644616 from the Simons Foundation. Resources supporting this work were provided by the NASA High-End Computing (HEC) Program through the NASA Advanced Supercomputing (NAS) Division at Ames Research Center.

\section*{Data Availability}
The data underlying this article will be shared on reasonable request to the corresponding author.



\bibliographystyle{mnras}
\bibliography{refs} 



\appendix

\section{Additional plots}

\begin{figure*}
    \centering
    \includegraphics[width=1.67\columnwidth]{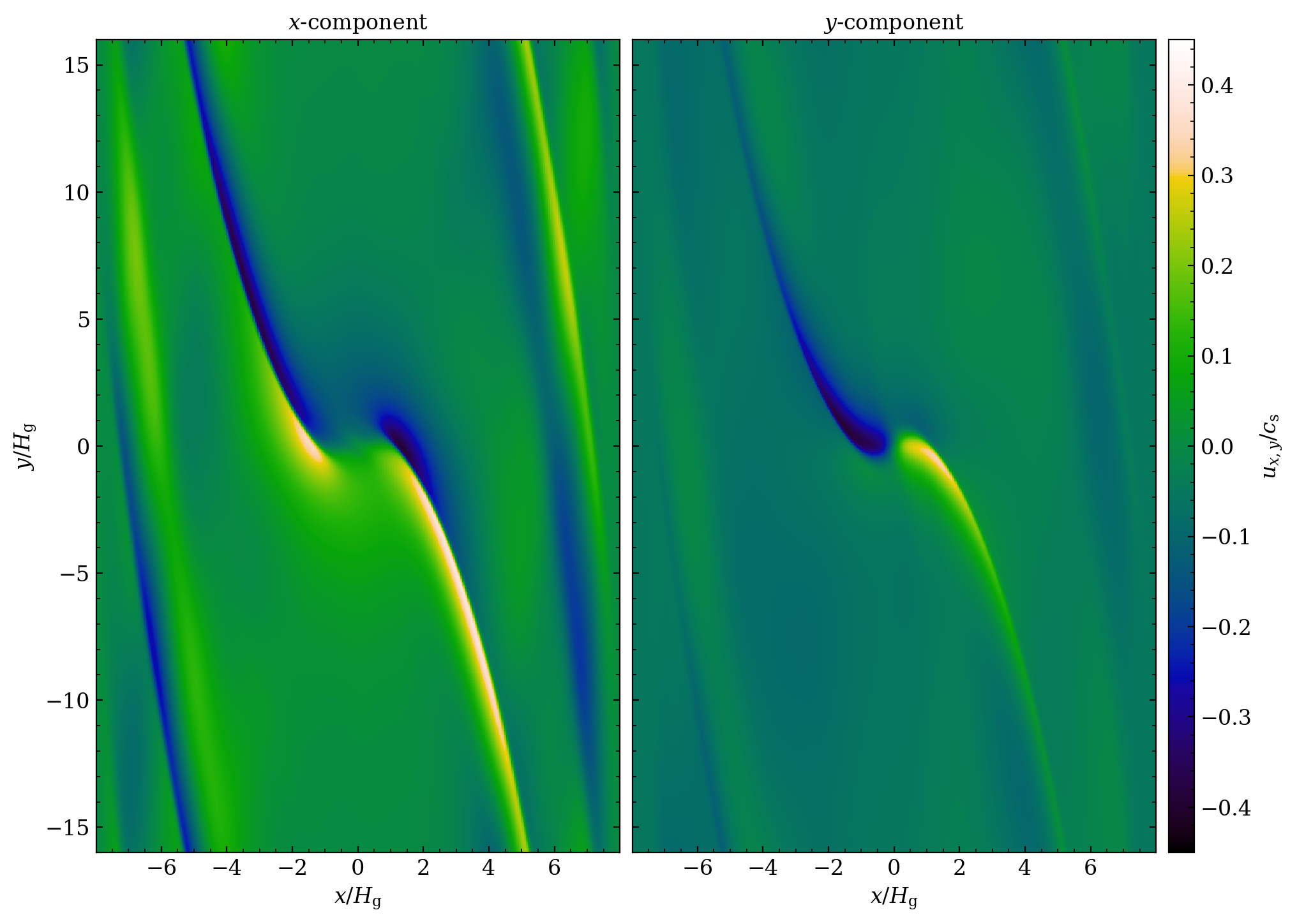}
    \caption{The x- and y-component of the gas velocity field at the equilibrium state from our simulation with $M_{\rm p}/M_{\rm th} = 0.5$. The background Keplerian shear has been removed from the velocity field.}
    \label{fig:yVSx_ux}
\end{figure*}

Fig.~\ref{fig:yVSx_ux} shows the x- and y-component of the gas velocity field at the equilibrium state from our simulation with $M_{\rm p}/M_{\rm th} = 0.5$. The velocity perturbations driven by the planet across spiral arms can be as high as $\sim$ 50\% the speed of sound. These perturbations are the likely source to drive high collision speeds between particles of different sizes within the spiral arms.

\bsp	
\label{lastpage}
\end{document}